  \def\versionno{ fixpoint -- version 12 -- by aw --  }

\def\be            {\begin{eqnarray}}
\def\chii          { \raisebox{.15em}{$\chi$} }
\def\complex       {{\ensuremath{\mathbb C}}}
\def\ee            {\end{eqnarray}}
\def\eE            {{\rm e}}
\newcommand\erf[1] {(\ref{#1})}

\newcommand\Frac[2]{\mbox{\large$\frac{#1}{#2}$}}
\def\ii            {{\rm i}}

\newcommand\query[1]{\ifnum\draftcontrol=1
                   \marginpar{{~}\\[-4em]\small label:\\``#1"\\[1em]~}\fi}

\newcommand\void[1]{}

\def\zet           {\ensuremath{\mathbb Z}}

\catcode`\@=11

\newif\if@fewtab\@fewtabtrue
{\count255=\time\divide\count255 by 60 \xdef\hourmin{\number\count255}
\multiply\count255 by-60\advance\count255 by\time
\xdef\hourmin{\hourmin:\ifnum\count255<10 0\fi\the\count255}}
\def\ps@draft{\let\@mkboth\@gobbletwo \def\@oddhead{}
    \def\@oddfoot{\hbox to 7 cm{\tiny \versionno
       \hfil}\hskip -7cm\hfil\rm\thepage \hfil {\tiny\draftdate}}
    \def\@evenhead{}\let\@evenfoot\@oddfoot}
\def\draftdate{\number\month/\number\day/\number\year\ \ \ \hourmin }
 \global\def\draftcontrol{0}

\def\draftcite#1{\ifnum\draftcontrol=1#1\else{}\fi}
\def\@lbibitem[#1]#2{\item{}\hskip -3\hbox to 2cm
{\hfil$\scriptstyle\draftcite{#2}$}\hskip
1cm[\@biblabel{#1}]\if@filesw {\def\protect##1{\string ##1\space}\immediate
      \write\@auxout{\string\bibcite{#2}{#1}}}\fi\ignorespaces}
\def\@bibitem#1{\item\hskip -3cm \hbox to 2cm
{\hfil {\footnotesize\draftcite{#1}}}\hskip 1cm
\if@filesw \immediate\write\@auxout
       {\string\bibcite{#1}{\the\value{\@listctr}}}\fi\ignorespaces}

\catcode`\@=12

\documentclass[12pt]{article}
\usepackage{latexsym,amsmath,amssymb,amsfonts,epsf,color,colordvi}
\usepackage[dvips]{graphics}
\usepackage[mathscr]{eucal}
\usepackage{rotating}
\numberwithin{equation}{section}

\def\hbar          {\ensuremath{{\liefont h}}}
\begin{document}

\begin{flushright}  {~} \\[-12mm] {\sf hep-th/0605056}\\[1mm]
 \end{flushright} \begin{center} \vskip 26mm
{\Large\bf Finite level geometry of fractional branes}\\[24mm]
{\large Albrecht Wurtz }\footnote{\tt albrecht.wurtz@kau.se}\\[12mm]
{\small
Karlstads Universitet\\
Universitetsgatan 5 \\
S\,--\,651\,88\, Karlstad} \\[5mm]

 \end{center} \vskip 28mm

\begin{quote}{\bf Abstract}\\[1mm]
In some CFT models of simple current type, which are used to 
describe string theory on orbifolds and (adjoint) cosets of Lie 
groups, there arise fixed points of the simple current group. In 
these cases, the standard procedure to associate functions to 
Ishibashi states by averaging out the action of the simple current 
group, gives functions with unsatisfactory properties. In some cases 
the averaged Ishibashi function simply vanishes, which we see 
explicitly in $SO(3)$ at level $k=4l{+}2$. In this note, an 
alternative function assignment is suggested, and it is shown that 
in some cases the resulting Ishibashi functions are orthogonal. 
\end{quote} \vfill \newpage

\section{Introduction} D-branes in string theory can be described in 
(at least) two different ways. One can, in the Lagrangian 
description, {impose} boundary conditions that restrict the end 
points of the string to live on certain subsets of the target space. 
Then, under certain conditions, the field theory on the world sheet 
is conformal \cite{Gaw,SSW}. Alternatively, one can introduce 
boundary states in CFT and {\it directly} impose conformal 
invariance and other symmetry conditions. These boundary conditions 
can then (less directly) be given a geometrical interpretation in 
terms of functions on the target space. This function is interpreted 
as describing the localization of the string endpoint. In free boson 
theories, the boundary states turn out to correspond in a natural 
way to delta functions on the target space, in perfect agreement 
with the imposed boundary conditions in the Lagrangian description 
\cite{DiV}. In WZW models, there is also an agreement between the 
descriptions, in the sense that the boundary states converge to 
certain delta functions in the limit of large level 
\cite{FFFS}\footnote{However, in the Lagrangian description, there 
are so-called exceptional branes with lower dimensionalities than 
the generic ones cf.\ \cite{Gaw}, unlike in the CFT description 
\cite{FFFS,fuwu}.}. One would like to generalize this procedure to 
sigma models on orbifolds of Lie groups,
and to coset models. \\[-2mm]

We consider boundary states that are rational, which are often 
symmetry preserving (they do not necessarily preserve the maximally 
extended symmetry). These boundary states can be used to construct a 
CFT for which correlators exist on all world sheets \cite{V}. In 
particular, they satisfy the Cardy constraints. The Ishibashi 
blocks, commonly referred to as Ishibashi states, are obtained as a 
natural basis of solutions of the Ward identities. In the WZW theory 
we start from (which we will refer to as the covering theory), the 
target space is a connected, simply connected compact Lie group $G$. 
The Peter-Weyl theorem provides an isomorphism between the functions 
on $G$ and the horizontal descendants of the primary fields. This 
isomorphism associates group characters to Ishibashi blocks 
\cite{FFFS}, which gives the boundary state a shape that converges, 
in the limit of large level, to a delta distribution on the group 
\cite{FFFS}. In CFT, the Ishibashi blocks are solutions to the Ward 
identities, thus it is natural that they are associated to group 
characters (that are invariant under $Ad_G$). Further, the 
characters are orthogonal with respect to the Haar measure on $G$, 
which reflects the corresponding orthogonality of the Ishibashi blocks.\\[-2mm]

 We 
consider the shape of boundary states in a target space $Q$ which is 
a $\Gamma$-orbifold of $G$ (we shall refer to $G$ as the covering of 
$Q$), where $\Gamma$ is a finite group. The Ishibashi blocks in the 
corresponding CFT are constructed with simple current methods 
\cite{FHSSW,V}. We would like to associate functions to these. A 
priori, it is not clear that this is at all possible. After all, in 
string theory, D-branes are more than just the sub-manifolds of 
target space where the string end-point is located. Here, our focus 
is only on those aspects of the D-branes which concern their 
position in target space. From this point of view, it would be 
preferable (if possible) 
to associate {\it functions} to Ishibashi blocks.  \\[-2mm]

 One procedure to associate functions to Ishibashi blocks
is to average out the action of $\Gamma$ on the Ishibashi functions 
on $G$, cf.\ \cite{MMS1,bowu}. (Or, equivalently, one averages the 
functions associated to the {\it boundary states} over the action of 
the orbifold group.) This gives functions that are manifestly 
invariant under the orbifold group action, and therefore descend to 
functions on the orbifold space $G/\Gamma$. 
 However, 
when there are fixed points, this procedure gives Ishibashi 
functions of which some are vanishing, or associates the same 
function to different Ishibashi blocks. This mismatch is 
unsatisfactory, because the Ishibashi blocks should be orthogonal 
\cite{FS1},
and it is natural to expect the same for the associated functions.  \\[-2mm]

The origin of this mismatch is the fact that in the CFT description, 
the Ishibashi blocks are not labeled with the same labels as the 
primary fields (to which functions are naturally associated). 
Instead we have pairs $(\Lambda_f,J)$, satisfying certain 
conditions, cf.\ \cite{huis,FHSSW}. Here, $\Lambda_f$ is a label of 
a primary field in the covering theory, $J$ is a simple current used 
to define the orbifold, and $J*\Lambda_f=\Lambda_f$. The symbol $*$ 
denotes the fusion product, under which the set of simple currents 
is a group.\\[-2mm]

For concreteness, we consider the orbifold $SU(2)/\zet_2=SO(3)$. 
There is a simple current $k$ and a fixed point $\frac{k}{2}$;  
$k*\frac{k}{2}=\frac k2$. We face two problems :\\[-2mm]

\noindent I) At level $k=4l+2$, the label $m=k/2$ occurs as (part 
of) an allowed Ishibashi label, namely $(k/2,k)$, but the function 
associated to this Ishibashi block by the averaging
procedure vanishes. \\
II) If instead $k=4l$, we have two allowed Ishibashi labels 
$(k/2,k)$ and $(k/2,0)$, which are both associated to the {\it same} 
function with the averaging procedure.\\[-2mm]

In both cases we run into trouble with the orthogonality of the 
Ishibashi blocks \cite{FS1} \be \langle \langle 
I_{m,J}|\,q^{L_0+\bar L_0-c/12}\, I_{n,K} \rangle \rangle = 
\delta_{nm}\delta_{JK}\chii_n(q^2)\,; \label{normalization} \ee 
functions which are linearly dependent cannot be orthogonal with any 
sensible scalar product. In the Lagrangian description, one 
associates the structure of a gerbe module to the branes \cite{Gaw}. 
Branes that are supported at the same sub-manifolds may differ by 
nothing more than their gerbe module structure, and this happens 
precisely when there are fixed 
points in the CFT description.\\[-2mm]

The main purpose of the present paper is to find a description that 
resolves the ambiguity II at finite values of the level. In section 
\ref{Ishi}, a procedure to associate functions to Ishibashi blocks 
is suggested for rank $r{=}1$, \erf{suggestion}. For higher rank 
groups, a universal procedure \erf{3-suggestion} is suggested and 
its implications are analyzed. The ansatz \erf{3-suggestion} is 
natural because it makes makes use of the isomorphism (see 
\cite{Yellow} eq. (14.110)) between the center $Z\subset G$ and the 
relevant simple current group $\mathcal G$. Further, it provides a 
universal procedure to associate functions to Ishibashi blocks such 
that the functions are (in many cases, including infinite series) 
orthogonal, as required. In the final part, we also propose a way to 
resolve the ambiguity I. In section \ref{SO3}, the implications of 
\erf{suggestion} are investigated. In the fourth and final section, 
the diagonal coset model $SU(2)_k\times SU(2)_l/SU(2)_{k+l}$ is 
discussed.

\section{Ishibashi functions}
\label{Ishi}

Choose a maximal torus $H\subset G$, and let $h\in \bar {\mathfrak 
h}^*=Lie(H)^*$. For $x \in \complex$, a representation function at 
$g\eE^{\ii x h}$ is related to its value at $g$ as \be 
D^{\Lambda}_{ab}(g\eE^{\ii x h})=\langle \Lambda;a | 
R^\Lambda(g\eE^{\ii x h}) |\,\Lambda;b \rangle= \eE^{\ii x(b ,h)} 
D^{\Lambda}_{ab}(g)\,, \ee where $a,b$ label states in the 
representation $\Lambda$ and $(\cdot,\cdot)$ is the Killing form 
(where we take the weight part of the label $b$, in case there are 
multiplicities). Recall \cite{BtD} that the functions 
$D^{\Lambda}_{ab}$ are orthogonal in all three indices with the 
usual scalar product (integration with the Haar measure). A 
character at this argument can be written as \be 
\chii_\Lambda(g\eE^{\ii x h})= \sum_{a\prec \Lambda}  \eE^{\ii x (a 
,h)} D^{\Lambda}_{aa}(g)\,. \ee The sum is over 
all states $a$ (with multiplicities) in the representation $\Lambda$.\\[-2mm] 

Let us, to begin with, focus on the case $SO(3)= SU(2)/\zet_2$. The 
two elements of the simple current group $\zet_2$ are denoted $0$ 
and $k$, and their action under fusion is $0*j=j$ and $k*j=k-j$. The 
simple current group has a fixed point under the simple current 
action; $k*k/2=k/2$. When $k=4l$, this fixed point corresponds to a 
primary field in $SO(3)$. Then we have Ishibashis which are 2-fold 
degenerate and are labeled by pairs $(\Lambda,J)$ with the 
degeneracy $J \in \{ 0,k\}= \{k^n\}$ with $n \in \{ 0,1 \}$ and 
$\Lambda = k/2$ is the fixed point.  Denote by $|\Lambda|$ the 
dimension of the (horizontal) $\mathfrak {su}(2)$-representation 
with highest weight $\Lambda$. One can associate to the Ishibashi 
labeled $(\Lambda,J)$ the function \be \langle g | 
I_{\Lambda,J}\rangle\rangle := \sqrt{\Frac{1}{|G|}}\,\, 
\chii_\Lambda(g\eE^{\ii \pi \sigma _3\frac{n}{|\Lambda|}})\,. 
\label{suggestion} \ee Here, $G=SU(2)$ and $|G|=2\pi^2(k\alpha 
')^{3/2}$ denotes the volume of the group \cite{BDS}. For both 
values of $n$, \erf{suggestion} defines a function on $SO(3)$. We 
shall see that they are not only linearly independent, but in fact 
orthogonal. Consider the scalar product of two Ishibashis with 
different labels $n=0$ and $n'=1$, calculated via their functions, 
\begin{eqnarray} \int_G {\rm dg}\,\langle \langle I_{ \Lambda,0 }| g 
\rangle \langle g |I_{ \Lambda, 1 }\rangle \rangle &=& 
\Frac{1}{|G|}\int_G {\rm dg} \sum_{a,b \prec \Lambda} \eE^{\ii\pi 
a/|\Lambda|} D^{\Lambda}_{aa}(g) D^{\Lambda}_{bb}(g)^*
\nonumber \\
&=&\frac{1}{|\Lambda|}\sum_{a \prec \Lambda}  \eE^{\ii\pi 
a/|\Lambda|} =0\,. \end{eqnarray} Recall $|\Lambda|=\Lambda {+}1$ 
and $a= {-}\Lambda,{-}\Lambda {+} 2,...,\Lambda{-}2,\Lambda$, so the 
sum is a sum over roots of unity $\eE^{\ii\pi a/|\Lambda|}$. 
Ishibashis with different degeneracy labels are now orthogonal, as 
they should be, see \cite{FS1} eq.\ (4.40). Moreover, the 
normalization of the individual Ishibashis is left unchanged. This 
is the prescription to resolve degenerate Ishibashi functions that 
will be applied in the coming section.\\[-2mm] 

The rest of this section is on generalizing the above procedure to 
higher rank groups. The simple currents of the WZW models (with one 
exception, which occurs for $E_8$) are labeled by $k$ times the 
fundamental weights $\Lambda_{(i)}$ with $a_i=1$  \cite{Fuchs}.  
These are the cominimal weights, cf. the table in \cite{Fuchs}, p 
203. The $\Lambda_{(i)}$ are the highest weights of the fundamental 
representations, and $\Lambda_{(0)}$ is the unit element in the 
fusion ring, hence also of the simple current group. $E_8$ at level 
$k=2$ has an additional simple current $\Lambda_{(7)}$. In all cases 
except $E_8$ at level $k=2$, there exists an isomorphism between the 
simple current group and the center of the group; $\mathcal G \cong 
Z(G)$, given by $g_{(J)}^\xi=\exp(\xi \ii H_{\Lambda_{(J)}})$ with 
$\xi\,{=}\,2\pi$, cf.\ \cite{Yellow}. Here the $H_{\Lambda_{(J)}}$ 
are appropriately normalized so that the exponential of a 
fundamental Weyl alcove is a fundamental domain of the $Ad_G$-action 
on $G$. When $\Lambda=\Lambda_f$ is a fixed point 
$\Lambda_{(i)}*\Lambda_f=\Lambda_f$, it makes sense to require 
$I_{(\Lambda_f,\Lambda_{(0)})}(g)=\chii_{\Lambda_f}(g)$. We also 
require that we have an action of the simple currents on the 
Ishibashis, 
$I_{(\Lambda_f,\Lambda_{(J)}*\Lambda_{(0)})}(g)=I_{(\Lambda_f,\Lambda_{(0)})}(gg_{(J)}^\xi)$,   
as in the case without fixed points \cite{fuwu}. In the light of the 
simple current - center isomorphism, the most general ansatz would 
appear to be 
$I_{(\Lambda_f,\Lambda_{(L)})}(g)=\chii_{\Lambda_f}(gg_{(L)}^{2\pi})$. 
However, as $g_{(L)}^{2\pi}\in Z(G)$, 
$\chii_{\Lambda}(gg_{(L)}^{2\pi})$ is a class function, which we 
know is a linear combination of characters. We require that the 
Ishibashi functions are linearly independent. For given $\Lambda$, 
there are $d_\Lambda^2$ orthogonal representation functions (by the 
theorem of Peter and Weyl, see \cite{BtD}). Guided by the 
isomorphism $\mathcal G \cong Z(G)$ as discussed above, we associate 
functions \be I_{(\Lambda_f,\Lambda_{(i)})}(g):= 
\sqrt{\Frac{1}{|G|}}\,\, \chii_{\Lambda_f} \Big(g\,\eE^{\ii 
H_{\Lambda_{(i)}} \xi }\Big)  \label{3-suggestion} \,, ~~~~~~ \xi 
\neq 2\pi\,.\ee to the Ishibashi blocks. We shall now investigate 
whether the prescription \erf{3-suggestion} leads to Ishibashi 
functions which are also orthogonal (with the right choice of 
$\xi$), in which case the basis for Ishibashis suggested in 
\cite{FS1} is indeed orthogonal. The scalar product between two 
Ishibashis with $\Lambda_f$ a fixed point, is \be \langle 
I_{(\Lambda_f,\Lambda_{(i)})}(g),I_{(\Lambda_f,\Lambda_{(j)})}(g) 
\rangle &=& \sum _{a\prec \Lambda ,\, b\prec \Lambda' } \eE^{\ii (a, 
\Lambda_{(i)})  \xi } \eE^{-\ii (b,\Lambda_{(j)}) \xi } \int_G dg\,  
D^{\Lambda_f}_{aa}(g) D^{\Lambda_f'}_{bb}(g)^*
 \nonumber\\ &=& 
\frac{1}{|\Lambda_f |}\, \,\bar \chii _{\Lambda_f}\big(\Lambda_{(i)} 
\xi-\Lambda_{(j)} \xi \big)\,.\ee Thus, the questions whether these 
Ishibashis are orthogonal reduces to a calculation of the 
corresponding horizontal Lie algebra character $\bar \chii 
_{\Lambda_f}\big(\Lambda_{(i)} 
\xi-\Lambda_{(j)} \xi \big)$.\\[-2mm]

When we consider the fixed point $\Lambda = m \rho$ (which appears 
in all Lie groups at level $k=mg^\vee$), it is useful to rewrite the 
Lie algebra character with the Weyl character formula \be \langle 
I_{(\Lambda,\Lambda_{(i)})}(g),I_{(\Lambda,\Lambda_{(j)})}(g) 
\rangle &=& \frac{1}{|\Lambda|}\,\frac{\sum_{\sigma\in W} |\sigma| 
\exp \left( \ii\xi\, \sigma(\Lambda{+}\rho), \Lambda_{(i)} 
{-}\Lambda_{(j)}  \right)}{\sum_{\sigma\in W} |\sigma| \exp 
\left(\ii \xi \,\sigma(\rho), \Lambda_{(i)}{-}\Lambda_{(j)}  
\right)} \ee If $\Lambda_{(i)}=\Lambda_{(j)}$, we have a character 
evaluated at the origin, which takes the value $|\Lambda|$. 
Otherwise, one can show that $\Lambda_{(i)}-\Lambda_{(j)}=\omega 
(\Lambda_{(k)})$ for some fundamental weight $\Lambda_{(k)}\neq 0$, 
and some Weyl group element $\omega \in W$. To prove this, we need 
to do a case by case analysis where \cite{KQ} is a useful reference. 
For illustration, consider $E_6$, which has a $\zet_3$ simple 
current group with cominimal weights $\Lambda_{(1)}$ and 
$\Lambda_{(5)}$. In terms of fundamental Weyl reflections 
$w_i(\lambda)=\lambda - \lambda ^i \alpha_{(i)}$ one finds (e.\ g.\ 
by an algorithm suggested in \cite{Capps}) that \be 
\Lambda_{(1)}-\Lambda_{(5)}= 
w_5\,w_4\,w_3\,w_2\,w_6\,w_3\,w_4\,w_5(\Lambda_{(5)})\,.\ee 
Similarly, one can show in all other cases as well that for all 
simple currents $\Lambda_{(i)},\Lambda_{(j)}$ the difference is a 
horizontal Weyl element acting on a fundamental weight 
$\Lambda_{(k)}\neq 0$.\\[-2mm] 

We use this fact to proceed with our calculation of the Ishibashi 
function scalar product, \be \langle 
I_{(\Lambda,\Lambda_{(i)})}(g),I_{(\Lambda,\Lambda_{(j)})}(g) 
\rangle&=& \frac{1}{|\Lambda|}\,\frac{\sum_{\sigma\in W} |\sigma| 
\exp \left[ \ii \xi (m{+}1) \left( \rho, \sigma (\Lambda_{(k)} ) 
\right)  \right]}{\sum_{\sigma\in W} |\sigma| \exp \left[ \ii \xi 
\left( \rho, \sigma (\Lambda_{(k)}) \right)  \right]}\ee We can now 
use the denominator identity to rewrite both the denominator and the 
numerator, \be  \langle 
I_{(\Lambda,\Lambda_{(i)})}(g),I_{(\Lambda,\Lambda_{(j)})}(g) 
\rangle&=& \frac{1}{|\Lambda|}\,\frac{\prod_{\alpha > 0}\sin \left( 
\Frac{1}{2}\xi(\alpha,(m{+}1)\Lambda_{(k)} ) \right) }{\prod_{\alpha 
> 0}\sin \left( \Frac{1}{2}\xi(\alpha,\Lambda_{(k)} ) \right)}\ee 
The product is over the positive roots. Some factors in both 
products vanish because $(\alpha^j,\Lambda_{(k)})=0$ for simple 
$\alpha^j$ with $j\neq k$. This happens at both sides of the 
fraction, thus we shall ignore those factors. The other factors 
where $\alpha$ contains $\alpha^k$ are the ones which do not vanish 
for all $\xi$. We shall choose $\xi$ such that \be \sin \left( 
\Frac{1}{2}(m{+}1)\xi(\alpha^k,\Lambda_{(k)} ) \right)=0~~~~{\rm 
and}~~~~\sin \left( \Frac{1}{2}\xi(\alpha^k,\Lambda_{(k)} ) 
\right)\neq 0\,.\ee Then the Ishibashis have the desired 
orthogonality. We set \be \xi \equiv \frac{2\pi 
}{(m{+}1)(\alpha^k,\Lambda_{(k)} )}\,. \label{xi}\ee All 
$(\alpha^k,\Lambda_{(k)} )$ are equal for those $\Lambda_{(k)}$ that 
appear in $\Lambda_{(i)}-\Lambda_{(j)}=\omega (\Lambda_{(k)})$, 
hence the Ishibashis are indeed orthogonal. Thus, for these Lie 
groups, our prescription yields a natural basis of Ishibashi 
functions that are orthogonal in the labels $m\rho,\Lambda_{(i)}$. 
These Ishibashi functions can then be used to build the boundary 
state functions, which describe the shape of the D-branes. This will 
be displayed in detail for $SU(2)/\zet_2= SO(3)$ below.\\[-2mm]

As an example of a fixed point that is not of the form $\Lambda = m 
\rho$, we consider $SU(4)$ at level $k=4$ where the simple current 
$\Lambda_{(2)}$ has the fixed point $\Lambda_f= 2\Lambda_{(0)}+ 2 
\Lambda_{(2)}$, the horizontal part of which is $\bar \Lambda = 
(0,2,0)$. The character is \be \bar \chii _ {(0,2,0)}(\ii \xi 
\Lambda_{(2)})&=& 10 + 2\cos(2\xi) + 8\cos(\xi)\,, \ee which does 
not vanish for any (real) value of $\xi$. As the simple current 
group has only two elements, and precisely one of the functions 
\erf{3-suggestion} associated to the fixed point Ishibashi labels is 
a class function, it is clear that the two functions are linearly 
independent. But to obtain an orthogonal basis of Ishibashis, it 
appears that we must take a linear combination of $\{ \Lambda_{(0)}, 
\Lambda_{(2)} \} $ to label the degeneracy. The simple current 
$\Lambda_{(2)}$ also has the fixed point $\Lambda_{f} = 
2\Lambda_{(1)}+ 2 \Lambda_{(3)}$; for this fixed point, the 
corresponding horizontal character does vanish with some choice of 
$\xi$. Thus, $\{ \Lambda_{(0)}, \Lambda_{(2)} \} $ provides in this 
case an orthogonal basis of degeneracy 
labels.\\[-2mm] 

That the Ishibashis can be made orthogonal for $\Lambda_{f} 
=(2,0,2)$, but not for $\Lambda_{f} =(0,2,0)$, is related to the 
fact that the former representation has more states (84 instead of 
20). Since the states in the representations are distributed on 
regular polygons, a large representation will typically have a 
smaller percentage of states orthogonal to the axis $\Lambda_{(2)}$. 
Thus, we expect the failure of the Ishibashi functions to be 
orthogonal in the $(0,2,0)$ case to be a small level phenomenon, 
which is supported by an analysis of other cases. For example, in  
$SU(6)$ at level $k=2$, the current $\Lambda_{(3)}$ has 3 fixed 
points of dimensions 20, 84 and 84. An analysis of the orthogonality 
of the Ishibashi functions \erf{3-suggestion} gives similar 
conclusions as in the case $SU(4)$ at level $k=4$.

\paragraph{The missing function} As already mentioned, in 
$SO(3)=SU(2)/\zet_2$ at level $k=4l+2$, we get a label $\kappa$ for 
the Ishibashi block which cannot in an obvious way be related to a 
unique function on the quotient space: Since $k/2$ is odd, the 
Ishibashi label $k/2$ does not naturally correspond to a 
representation of the Lie group $SO(3)$. However, we can take a 
fundamental domain (of the $\Gamma$-action) $D\subset M$ in the 
covering space $M$ (which in our applications is a Lie group, but 
the considerations are general).  On $M$, we {\it do} have a natural 
Ishibashi function $I^M_\kappa$ for that label. Now we can take the 
restriction $I^M_\kappa |_D$, and extend it to a function $\tilde 
I_\kappa$ on $M$ that is invariant under the orbifold group $\Gamma$ 
(and which is different from the function $I^M_\kappa$). This 
function projects to a candidate for the Ishibashi function 
$I^Q_\kappa$ on the quotient $Q=M/\Gamma$. Since the integration 
$\langle I_\kappa , I_\Lambda \rangle_Q$ can be lifted to $D$, and 
if $\langle I_\kappa , I_\Lambda \rangle_D=0$ for $\kappa \neq 
\Lambda$, the Ishibashi functions are orthogonal. We shall spell out 
explicitly how this works for $M= SU(2)$ and $\Gamma =\zet_2$ in the 
next section, and we shall see that the suggested function is indeed 
orthogonal to the other Ishibashi functions on $SO(3)$.


\section{Ishibashis of SO(3)}
\label{SO3}

The CFT $SO(3)_{k}$ can be described as a simple current extension 
of $SU(2)_{k}$ when $k=4l$, and as a permutation invariant when 
$k=4l+2$. The primary fields in the covering $SU(2)_{k}$ are labeled 
$j=0,1,...,k$. The simple current group is $\mathcal G = 
\zet_2=\{k,0 \}$. The nontrivial element acts via fusion as 
$k*j=k-j$ and its fixed point is $k/2$. The Ishibashis are labeled 
$(m,J)$ with \be m=J*m \,,~~~~~~ Q_{\mathcal G}(m)+ X(\mathcal G, J) 
\in \zet \,, \label{allowed} \ee cf.\ \cite{FHSSW}. Since $\mathcal 
G$ is cyclic, the discrete torsion is trivial and $X$ vanishes 
except for $ X(k,k)=\frac{1}{2} \,Q_k(k)=  \frac{k}{4}$ mod $\zet$. 
When $k=4l+2$, we have discrete torsion in the sense $X \neq 0$ (but 
not in the sense of a freedom of choice in $X$). There are two types 
of Ishibashi labels in the simple current construction; the regular 
type \be (m,0)~~~~ {\rm with}~~~~m \in 2\zet\,, \label{list} \ee and 
one Ishibashi of exceptional type \be (\Frac{k}{2},k) ~~~~ {\rm 
with}~~~~ k\in 2\zet \,. \label{liste} \ee Both species are always 
present, and the first type Ishibashis are labeled by the allowed 
primary fields. The Ishibashis are expected to be orthogonal to each 
other, \cite{FS1}, and therefore the function cannot only depend on 
more the first entry in the pair $(m,J)$. Thus, the associated 
function cannot simply be the character $\chii_m(g)$ averaged over 
the orbifold group; instead \erf{suggestion} is suggested. 
Alternatively, we take \erf{3-suggestion} with $\xi$ given by 
\erf{xi}, which reduces to \erf{suggestion} in this case. Recall 
that the group manifold $ SO(3) =SU(2)/\zet_2 $ 
    is the set of equivalence classes
\be [g]=[zg]~~~~~~g\in SU(2)~~~~~~z=-\mathbf 1=  \eE^{\ii \pi \sigma 
_3}\,. \ee The stabilizer of this identification is trivial for all 
$g$. A function on $SO(3)$ is a function on the covering for which 
$f(zg) = f(g)$. In the case $k=4l+2$ we have \be 
D^{k/2}_{mm}(zg)=-D^{k/2}_{mm}(g)\,. \ee Simply averaging a sum of 
these over the orbifold group gives an everywhere vanishing 
function. Therefore, the group character $\chii_{k/2}(g)$ cannot be 
projected to $SO(3)$ by this averaging procedure. Instead, we may 
proceed as follows: for each $[g]\in SO(3)$ pick a representative 
$g$ that lies in the upper hemisphere of $SU(2)$, which is the set 
of points with $\psi < \pi/2$ in the parametrization \be g=\cos \psi 
\, \mathbf 1 +\sin \psi \,\bar \sigma\cdot \bar n\,.\ee
 The character $\chii_{k/2}$ (which is an odd function of $\psi$ on
$SU(2)$), can be modified to an even function $\tilde \chii_{k/2}$
on $SU(2)$ by defining $\tilde \chii_{k/2} := \chii_{k/2} $ on the
upper hemisphere, and $\tilde \chii_{k/2} := -\chii_{k/2} $ on the
lower hemisphere. Note that $\chii_{k/2} = 0 $ on the equatorial
plane. To the Ishibashi labelled $(k/2,k)=(2l+1,k)$, we associate
\be I_{(2l+1,k)}([g]) = \sqrt{\Frac{1}{|G|}} \,\, \tilde \chii_{k/2} (g) 
\label{JF-sugg} \,.\ee This function is orthogonal to all 
$SO(3)$-characters: the allowed $SO(3)$-characters are of the form 
$\chii_{2n}$ and are even functions as well, thus we may evaluate 
the scalar product by just integrating over the upper hemisphere. 
Then we can use invariance of the measure to obtain \be  \langle 
\tilde \chii_{k/2},\chii_{2l}\rangle \nonumber = \int_0^{\pi/2} 
d\mu_\psi \, \chii_{k/2}(\psi)\chii_{2l}(\psi) - \int_{\pi/2}^{\pi} 
d\mu_\psi \, \chii_{k/2}(\psi)\chii_{2l}(\psi) =0 \,. \ee A similar 
calculation reveals that $\langle \tilde \chii_{k/2},\tilde 
\chii_{k/2}\rangle =\langle \chii_{k/2},\chii_{k/2}\rangle $.

\paragraph{Boundary states} Recall that the boundary labels are 
orbits $[j,\psi_j]$, where $j$ is an $SU(2)$ label that representats 
a $\mathcal G\cong \zet _2$ orbit, and $\psi_j$ is a character of 
$C_j\subset S_j$. The subgroup $C_j$ of the stabilizer $S_j$ is in 
our case given by $C_j=S_j$. Further, $C_{k/2}=\zet_2$, and all 
other $C_j$ are trivial. Hence $\psi_j$ is a degeneracy label that 
takes values $\pm 1$. A list of boundary labels is (supressing 
trivial labels) \be [j,\psi_j]= 
[0],\,[1],...,[k/2{-}1],\,[k/2,1],\,[k/2,{-}1]\,. \ee The boundarry 
states $[j]$ with odd $j$ do not correspond to primary fields of 
$SO(3)_{2k}$ and are interpreted as being symmetry-breaking. There 
are $k/2+2$ states in this list, just as many as there are 
Ishibashis \erf{list} and \erf{liste}. The boundary sates are linear 
combinations of the characters with coefficients \cite{FHSSW} \be 
B_{(m,J),[j,\psi_j]}= \sqrt{\frac{|\mathcal G|}{|S_j||C_j|}}\, 
\frac{\alpha_J S^J_{m,j}}{\sqrt{S_{0,m}}}\psi_j(J)^* =\frac{\sqrt 
2}{|S_j|}\frac{\alpha_J S^J_{m,j}}{\sqrt{S_{0,m}}}\psi_j(J)^*\,, \ee 
which are known as the boundary coefficients. The matrix $S^J$ is 
different from the modular matrix $S$ only if $m=j=k/2$. In that 
case, $S^k_{k/2,k/2}=\Frac{1}{k}\,\eE^{-3\pi \ii k/8}$. The phase 
$\alpha_J$ can be taken to be $\alpha_k=\eE^{\ii \pi /4}$ when 
$k=4l+2$ and unity in all other cases.

\paragraph{Shape of the boundary states} Now we are ready to compute 
some of these boundary shapes. At level $k=4l$, we have a fractional 
boundary state with $j=k/2=2l$. With \erf{suggestion}, the shape of 
the fractional boundary state is  \begin{eqnarray} B_{[2l,\psi]}(g) 
&=& \frac{1}{\sqrt 2 } \sum _{(m,J)} 
\frac{S^J_{m,2l}}{\sqrt{S_{0,m}}} \psi(J) I_{(m,J)}(g)
\nonumber \\
&=& \frac{1}{\sqrt 2 } \sum _{m=0,2,...,k}  
\frac{S_{m,2l}}{\sqrt{S_{0,m}}} \chii_{m}(g) +\psi(k) \frac{1}{4l 
\sqrt 2 }  \frac{\eE^{-3\pi \ii l/2}}{\sqrt{S_{0,2l}}} 
\chii_{2l}(\eE^{\frac{\ii \pi}{2l+1}\sigma_3 }  g)\,, \end{eqnarray} 
where $\psi(k)=\pm 1$. The shift $g\mapsto \eE^{\frac{\ii 
\pi}{2l+1}\sigma_3 }  g$ is interpreted as a tilting of the 
conjugacy class by an angle $\pi/(2l+1)$. In \cite{bowu} and 
\cite{MMS1} the $\psi$-dependent last term does not contribute with 
a full group character.


\section{The coset [SU(2)$\times$SU(2)]/Ad(SU(2)) }

The prescription \erf{suggestion} can be applied to coset models 
with nontrivial field identification fixed points, in particular to 
the coset \be Q= \frac{SU(2)\times SU(2)}{{\rm Ad}(SU(2))} \, , 
\label{pillow} \ee which is the set of equivalence classes $ 
[g_1,g_2]=[gg_1g^{-1},gg_2g^{-1}]$ with $g,g_1,g_2\in G$. We use the 
parametrization \be Q = \left\{ \,(t,q) ~\, \bigg| \,~ t\in 
T_W\,,\,~ \begin{array}{c}
            q\in T_W  ~~ {\rm if}~~t=\pm e        \\[2mm]
            q\in G/{\rm Ad}(T) ~~ {\rm else}
           \end{array}
\right\}\,. \ee The CFT description of a sigma model with target 
space an adjoint coset $Q=G/{\rm Ad}(H)$, where $H \subset G$, goes 
as follows. The primary fields are labelled by certain pairs $(j,m)$ 
where $j$ is a primary in $G$ and $m$ a primary in $\bar H$. Here 
$\bar H$ is a certain CFT related to $H$ (the modular tensor 
category describing the primary fields of $\bar H$ is the dual of 
the MTC associated to $H$ \cite{correspondences}). Not all pairs 
$(j,m)$ of labels appear in the coset, there are selection rules and 
field identifications, which (for non-Maverick cosets) can be 
obtained by a simple current construction. The relevant simple 
current group (in this setting called identification group) is the 
one that geometrically corresponds to $Z_G \cap H$. In our model 
\erf{pillow}, the identification group is generated by $(-e,-e)\in 
G\times G$, which corresponds to $\zet_2 = \{(0,0,0),(k,l,k+l)\}$. 
The monodromy of a label with respect to this current is required to 
be integer, which leads to the selection rule \be a+b-c \in 2 
\zet\,. \label{ds}\ee The identification group has a fixed point if 
both levels $k$ and $l$ are even, which 
then is $(k/2,l/2,k/2+l/2)$. \\[-2mm]

Now we wish to describe the shape of the boundary states in this 
model. Most of the analysis is similar as in the case without fixed 
points (cf.\ \cite{fuwu}), only that we need something like 
\erf{suggestion} to associate functions to the degenerate 
Ishibashis. One can find the Ishibashi and boundary labels by 
following the standard procedure. The self- monodromy of the 
non-trivial identification current is two times its conformal 
weight, which is always integer. Hence the discrete torsion $X$ 
vanishes. Therefore, the Ishibashis are labeled by tuples $(m,J)$ 
where the monodromy of $m$ with respect to $J$ vanishes, and 
$J*m=m$. Thus, the first species of Ishibashis are simply the 
primary fields allowed by \erf{ds}. For even levels $k,l\in 2\zet $ 
there will be an additional fractional Ishibashi labeled \be 
(m;K)=(k/2,l/2,k/2{+}l/2\,;\,k,l,k{+}l)\,. \ee In order to associate 
functions to primary fields, it is convenient to first rewrite the 
(adjoint) coset \erf{pillow} as a set of equivalence classes 
\cite{fuwu,FrS} \be [g,h]_{lr} \, = \, [ugv,uhv]_{lr}\,, ~~~~~~~ u,v 
\in H \,. \ee To the Ishibashis labeled $(a,b,c;0)$ we associate a 
function which is invariant under the coset action; \be 
I_{(a,b,c,0)}[ug_1v,ug_2v,uhv] =I_{(a,b,c,0)}[g_1,g_2,h]\,, \ee 
which is achieved by the function \cite{fuwu}, eq.\ 
(5.7)\footnote{There is a small error in that formula: the Ishibashi 
function should be {\it without} the averaging over the simple 
current group. This distinction is insignificant in the context of 
\cite{fuwu}, where the focus is on boundary state functions; the sum 
over the identification group comes in in the next step, so the 
boundary functions obtained in \cite{fuwu} are still correct.}. 
Applied to the present context, the formula in \cite{fuwu} involves 
the Clebsch-Gordan coefficients of $SU(2)$. In our notation $c$ is 
contained in $a \times b$ if $c= |a{-}b|,...,a{+}b$, and the 
magnetic quantum number $\gamma$ is in the representation $c$ if 
$|\gamma|\leq c$, with multiplicity given by the Clebsch-Gordan 
coefficient $c^{c\prec a\times b}_{\gamma,\alpha,\beta}$. Recall 
that the Clebsch-Gordan coefficients are defined by \be c^{j \prec 
j_1{\times}j_2}_{m,m_1,m_2} = \langle j,m|\Big( |j_1,m_1\rangle 
\otimes |j_2,m_2\rangle \Big)\,. \label{CG} \ee Note that $c^{c\prec 
a\times b}_{\gamma,\alpha,\beta}=0$ if $\gamma \neq \alpha+\beta$. 
Also note that $\sum_{\alpha,\beta,\gamma} |c^{c\prec a\times 
b}_{\gamma,\alpha,\beta}|^2 = d_a d_b = d_c $. The regular Ishibashi 
function after integrating out the coset action is \be \label{oishi} 
I^Q_{a,b,c}[g_1,g_2,h] =\sqrt{\frac{|G_{k+l}|^5}{|G_{k}||G_{l}|}} 
\sqrt{\frac{d_ad_b}{d_c^3}} \sum_{\delta,\epsilon \prec a\atop 
{\mu,\nu \prec b }} D^a_{\delta \epsilon}(g_1) D^b_{\mu \nu}(g_2) 
D^c_{\delta + \mu , \epsilon+\nu}(h)^* \big( c^{c\prec a\otimes 
b}_{\delta+\mu,\delta,\mu}  \big)^* c^{c\prec a\otimes 
b}_{\epsilon+\nu,\epsilon,\nu}\,. \ee The shape of the fractional 
Ishibashi is the analoguous projection of the product of twisted 
characters, \be \label{Kishi} &&I^Q_{K}[g_1,g_2,h] \, = \, 
\sqrt{\frac{|G_{k+l}|^5}{|G_{k}||G_{l}|}} 
\,\sqrt{\frac{d_ad_b}{d_c^3}}\,
\\[1mm] \nonumber
&& ~~~~~~~~~~~ \times \sum_{\delta,\epsilon \prec a\atop {\mu,\nu 
\prec b }} D^{k/2}_{\delta \epsilon}(g_1\eE^{\ii \sigma_3 2\pi /k}) 
D^{l/2}_{\mu \nu}(g_2\eE^{\ii \sigma_3 2\pi /l}) 
D^{(k+l)/2}_{\delta+\mu, \nu+\epsilon}(h\eE^{\ii \sigma_3 2\pi 
/(k+l)})^* \big( c^{c\prec a\otimes b}_{\delta+\mu,\delta,\mu}  
\big)^* c^{c\prec a\otimes b}_{\epsilon+\nu,\epsilon,\nu} \,. \ee 
The Ishibashi function \erf{Kishi} is orthogonal to the other 
Ishibashi functions \erf{oishi}, because the scalar product can be 
lifted to the covering $G\times G \times G$ where orthogonality 
follows from previous considerations. The shapes of the boundary 
states can now be calculated by using the boundary coefficients 
given in \cite{FHSSW}. The results of this calculation are more
complicated than enlightening; we refrain from presenting them here.

\paragraph{Acknowledgements} The author would like to thank J\"urgen Fuchs for helpful comments.


\begin{thebibliography}{99} \bibitem{Gaw} K. Gawedzki, {\it Abelian 
and non-Abelian branes in the WZW models and gerbes}, Commun. Math. 
Phys. 258 (2005) 23-73, {\tt http://xxx.lanl.gov/abs/hep-th/0406072} 

\bibitem{SSW} U. Schreiber, C. Schweigert and K. Waldorf, {\it 
Unoriented WZW Models and Holonomy of Bundle Gerbes}, ZMP-HH/05-28, 
Hamburger Beitr. zur Mathematik Nr. 228 {\tt 
http://xxx.lanl.gov/abs/hep-th/0512283} 

\bibitem{DiV} P.\ Di\ Vecchia, M.\ Frau, I.\ Pesando, S.\ Sciuto, 
A.\ Lerda, R.\ Russo,\\ {\it Classical p-branes from boundary 
state}, Nucl. Phys. B 507 (1997) 259-276\\ {\tt 
http://xxx.lanl.gov/abs/hep-th/9707068} 

\bibitem{FFFS} G. Felder, J. Fr\"ohlich, J. Fuchs and C. Schweigert 
{\it The geometry of WZW branes}, J. Geom. Phys. 34 (2000) 162-190 
{\tt http://xxx.lanl.gov/abs/hep-th/9909030}

\bibitem{fuwu} J. Fuchs and A. Wurtz, {\it On the geometry of coset 
branes}, Nucl.Phys. B724 (2005) 503-528, {\tt 
http://xxx.lanl.gov/abs/hep-th/0505117}

\bibitem{V} J.\ Fjelstad, J.\ Fuchs, I.\ Runkel, C.\ Schweigert, 
{\it TFT construction of RCFT correlators V:\\ Proof of modular 
invariance and factorisation}, to appear in Theory and Applications 
of Categories, {\tt http://xxx.lanl.gov/abs/hep-th/0503194}


\bibitem{FHSSW} J. Fuchs, L. Huiszoon, A. Schellekens, C. Schweigert 
and J. Walcher, {\it Boundary coefficients and simple currents and 
Ishibashi labels}, Phys.Lett. B495 (2000) 427-434 {\tt 
http://xxx.lanl.gov/abs/hep-th/0007174 }


\bibitem{MMS1} J. Maldacena, G. Moore and N. Seiberg, {\it 
Geometrical interpretation of D-branes in gauged WZW models}, JHEP 
0107 (2001) 046, {\tt http://xxx.lanl.gov/abs/hep-th/0105038} 

\bibitem{bowu} P. Bordalo and A. Wurtz, {\it D-branes in lens 
spaces}, Phys. Lett. B 568 (2003) 270-280 {\tt 
http://xxx.lanl.gov/abs/hep-th/0303231}

\bibitem{FS1} J. Fuchs, C. Schweigert, {\it Symmetry breaking 
boundaries I, General theory}, Nucl.Phys. B558 (1999) 419-483, {\tt 
http://xxx.lanl.gov/abs/hep-th/9902132}

\bibitem{huis} L. Huiszoon, {\it D-branes and O-planes in string 
theory}, PhD thesis, Amsterdam 2002. 



\bibitem{BtD} T. Br\"ocker, T. tom Dieck, {\it Representaions of 
compact Lie groups}, Springer 1985

\bibitem{BDS} C.\ Bachas, M.\ Douglas, C.\ Schweigert {\it Flux 
stabilization of D-branes},\\ JHEP 0005 (2000) 339, {\tt 
http://xxx.lanl.gov/abs/hep-th/0003037}

\bibitem{Fuchs} J. Fuchs, {\it Affine Lie algebras and quantum 
groups}, Cambridge University Press, 1992 

\bibitem{Yellow} P.\ Di Franscesco, P.\ Mathieu, D.\ Senechal, {\it 
Conformal Field Theory}, Springer 1997

\bibitem{KQ} R.\ King, A. Al-Qubanchi, {\it The Weyl groups and 
weight multiplicities of the exceptional Lie groups}, J.\ Phys.\ A: 
Math.\ Gen.\ 14 (1981) 51-75



\bibitem{Capps} R. Capps, {\it The Weyl orbits of $G_2$, $F_4$, 
$E_6$ and $E_7$},\\ J.\ Phys.\ A:\ Math.\ Gen.\ 22 (1989) 1223-1243

\bibitem{correspondences} J. Fr\"ohlich, J. Fuchs, I. Runkel, C. 
Schweigert, {\it Correspondences of ribbon categories}, Adv. Math. 
199 (2006) 192-329, {\tt http://xxx.lanl.gov/abs/math.CT/0309465}

\bibitem{FrS} S. Fredenhagen and V. Schomerus, {\it D-branes in 
coset models}, JHEP 0202 (2002) 005, {\tt 
http://xxx.lanl.gov/abs/hep-th/0111189} 


























\end{thebibliography}
 \end{document}